\documentstyle[12pt,epsfig,graphics,color,amsmath,amssymb,latexsym]{article}
\def\baselinestretch{1.3}
\newcommand{\lapprox}{%
\mathrel{
\raise0.6ex\copy0\kern-\wd0
\lower0.65ex\hbox{$\sim$}}}
\newcommand{\gapprox}{%
\mathrel{
\raise0.6ex\copy0\kern-\wd0
\lower0.65ex\hbox{$\sim$}}}
\newcommand{\ba}{\begin{array}}
\newcommand{\ea}{\end{array}}
\newcommand{\bd}{\begin{displaymath}}
\newcommand{\ed}{\end{displaymath}}
\newcommand{\be}{\begin{equation}}
\newcommand{\ee}{\end{equation}}
\newcommand{\bea}{\begin{eqnarray}}
\newcommand{\eea}{\end{eqnarray}}
\newcommand{\tc}{\textcolor}
\newcommand{\gr}{\textcolor{green}}
\newcommand{\bl}{\textcolor{blue}}
\newcommand{\rd}{\textcolor{red}}
\newcommand{\ye}{\textcolor{blue}}
\newcommand{\pk}{\textcolor{pink}}
\def\barr{\begin{array}}
\def\earr{\end{array}}
\def\dis{\displaystyle}
\def\bra{\langle}
\def\ket{\rangle}
\def\a{\alpha}
\def\as {\alpha_s}
\def\b{\beta}
\def\g{\gamma}
\def\d{\delta}
\def\e{\epsilon}
\def\ve{\varepsilon}
\def\l{\lambda}
\def\m{\mu}
\def\n{\nu}
\def\G{\Gamma}
\def\D{\Delta}
\def\L{\Lambda}
\def\s{\sigma}
\def\p{\pi}
\def\stop{\tilde t}
\def\mzs {M_Z^2}
\def\mws {M_W^2}
\def\q2 {q^2}
\def\sz {\sin^2\theta_W}
\def\cz {\cos^2\theta_W}
\def\r {\rightarrow}
\def\t {\times }
\def\miss {\hspace{-0.5cm}\slash~~}
\def\photino {\tilde\gamma}
\def\neu {\chi_1^0}
\def\rslep {\tilde{e_R}}
\def\rsnu {\tilde{\nu}_R}
\def\lsnu {\tilde{\nu}_L}
\def\snu {\tilde{\nu}}
\def\lslep {\tilde{e_L}}
\def\stau {\tilde{\tau}}
\def\mer {m_{\rslep}}
\def\mmr {m_{\tilde{\mu}_R}}
\def\mml {m_{\tilde{\mu}_L}}
\def\mel {m_{\lslep}}
\def\mneu {m_{\neu}}
\def\bt{\begin{table}}
\def\et{\end{table}}
\catcode`@=11 
\def \gsim{\mathrel{\mathpalette\@versim>}}
\def \lsim{\mathrel{\mathpalette\@versim<}}
\def \@versim#1#2{\lower0.4ex\vbox{\baselineskip\z@skip\lineskip\z@skip
\lineskiplimit\z@\ialign{$\m@th#1\hfil##\hfil$%
\crcr#2\crcr\sim\crcr}}}
\catcode`@=12 

\begin{document}
\begin{titlepage}   
\begin{flushright}
{HRI-P 08-02-002\\HRI-RECAPP-08-01}
\end{flushright}
\vspace*{0.2cm}
\begin{center}
{\Large {\bf Right sneutrinos and signals of a stable stop at the Large Hadron Collider}}\\[0.3cm]

Debajyoti Choudhury$^\dagger$\footnote{\tt E-mail:debajyoti.choudhury@gmail.com},
Sudhir Kumar Gupta$^\star$\footnote{\tt E-mail:guptask@mri.ernet.in} \\
and Biswarup Mukhopadhyaya$^\star$\footnote{\tt E-mail:biswarup@mri.ernet.in}\\[0.3cm]
{\em $^\dagger$\hspace{-.2cm} Department of Physics and Astrophysics\\
University of Delhi, Delhi--100 007, India\\[.2cm]
$^\star$\hspace{-.2cm} Regional Centre for Accelerator-based Particle Physics \\
\it Harish-Chandra Research Institute \\
Chhatnag Road, Jhusi, Allahabad--211 019, India}\\[.2cm]
\end{center}

\begin{abstract} We investigate charged tracks signals of a supersymmetric
scenario, where the lighter stop is the next-to-lightest supersymmetric
particle (NLSP). It is found that such an NLSP is stable on the scale of
the detector at the LHC if one has a right-chiral sneutrino as the
lightest supersymmetric particle (LSP). After identifying some benchmark
points in the parameter space of a supergravity scenario with
non-universal scalar masses, we study a few specific classes of signals,
namely, stop pair production and gluino pair production followed by each
decaying into a stop and a top. It is shown that proper kinematic cuts
remove the backgrounds in each case, and, an integrated luminosity of even
1 $fb^{-1}$ is likely to yield copious events of the first kind, while a
larger lumionsity may be required for the other type. One can also
aspire to reconstruct the gluino mass, using the `visible' stable NLSP
tracks. 
\end{abstract} 
\end{titlepage} 
\newpage 
\setcounter{footnote}{0}
\def\baselinestretch{1.5}

\section{Introduction}

    Among the various new physics possibilities at the TeV scale,
supersymmetry (SUSY) \cite{susyrev} occupies a slightly preferred
position. From a bottom-up point of view, one reason behind this is
the dual role of SUSY in stabilizing the electroweak (EW) scale and,
in its minimal version, in offering a cold dark matter candidate in
the form of the lightest supersymmetric particle (LSP). 
From the top-down standpoint, too, SUSY broken at
the TeV scale fits in rather well in scenarios such as supergravity
(SUGRA), which, presumably, have a close connection to physics 
at the Planck scale.
Therefore, despite some persistent concerns such as the possible
enhancement of flavour-changing neutral currents (FCNC), one feels the
urge to fit in any proposed SUSY scenario into a scheme where the SUSY
breaking parameters evolve down from values inherited at a high
scale. Although the simplest model to achieve this is the minimal
SUGRA (mSUGRA) picture, scenarios with non-universal masses at high
scales are also often viable.

Indeed, one has to go beyond the minimal version (of the standard
model as well as its SUSY extension) if one has to explain the
accumulating evidence in favour of neutrino masses and mixing 
\cite{revnu}. The simplest way to do this is to postulate a
right-handed neutrino in each generation. In a SUSY version, this
entails right-chiral sneutrinos \cite{rneut}.  While the minimal SUSY
standard model (MSSM) favours the lightest neutralino as the LSP,
right-sneutrino LSP's are equally viable if the particle content is
extended in the manner suggested above.  This is particularly true if
the neutrinos have only Dirac masses \cite{moroi}, for the existence of
$\Delta L = 2$ terms in the Lagrangian nominally leading to large
(keeping in view the seesaw mechanism) Majorana masses would
simultaneously elevate the right-sneutrino masses to high
values\footnote{A possible exception to this rule may be provided by
situations wherein the right-handed neutrino mass matrix has a
vanishing determinant, occasioned, for example, by texture zeroes.
While such scenarios may arise naturally in models with an extended
symmetry texture and may lead to very interesting phenomenology, we
refrain from discussing those here.}. A right sneutrino LSP evades the
limits from direct dark matter search due to its near-sterile
character. Its viability as cold dark matter candidate~\cite{moroi} 
has also been demonstrated, although there are debates about the possible
non-thermal nature~\cite{c8thermal} of such dark matter.

If neutrinos have only Dirac masses, then the interactions of an LSP
dominated by right sneutrinos would be proportional to the neutrino
Yukawa couplings $y_{\nu}$, which are of order $10^{-12}$ or
less. This is because (a) if it is a scalar trilinear interaction,
then it is proportional to $y_{\nu}$, and (b) if it is gauge
interaction, then it is proportional to the overlap of the LSP with
left sneutrino, which, by virtue of the left-right mixing terms in
sfermion mass matrices, is again proportional to the neutrino mass.
Thus, the decay of the next-to-lightest supersymmetric particle (NLSP) 
to the LSP takes place over lengths much 
larger than the scale of collider detectors, and SUSY signals are
drastically different from those of MSSM where missing transverse
momentum is the key distinguishing feature~\cite{drees}. Two of us 
have shown, in an
earlier work, how, in such cases, a stau-NLSP can provide signals that
can be distinguished from the standard model (SM) background
\cite{stau}.  In this work, we discuss the signals characteristic of a
quasi-stable stop NLSP.

Needless to say, right sneutrinos make a big difference to the signal
if the NLSP is a charged particle, which would leave tracks in the inner
tracker as well as the muon chamber. Apart from the stau (or, in
special cases, sleptons of the first two families), a possible NLSP is
either a chargino or a lighter squark of the third family. 
Theoretically though, it is difficult to render a chargino 
the lightest of the SM's supersymmetric partners; in other words, 
the coexistence of a chargino NLSP with a right-sneutrino LSP 
is very difficult to accommodate. even if we assume non-universal gaugino 
mass. A stop NLSP with a right-sneutrino LSP,
on the other hand, can lead to very interesting signals at the Large
Hadron Collider (LHC). Cases where such a stop decays within the
detector have been studied in an earlier work \cite{gopal, yudi}. We
think that it is equally interesting to consider the situation where
the stop NLSP decays into the right-sneutrino LSP through the
sneutrino Yukawa coupling and, thus, escapes the muon chamber after
leaving a track there. We show that, just like the case with a stau
NLSP, kinematical separation of such signals from the SM backgrounds
\cite{bkg} is clearly possible, making such SUSY scenarios eminently
distinct. Moreover, it is also possible to distinguish a stop-NLSP
scenario from one with a stau NLSP, 
simply from a comparison of stop-pair production and stau-pair production
(in the alternate scenario). Furthermore, the study of
 stops as intermediates in gluino cascade decays provides 
additional discriminants.

Unlike the case of a stau NLSP, a stop NLSP is difficult to obtain in
a SUGRA setting with universal scalar masses at the high scale. On the
other hand, such a spectrum can arise naturally when the scalar masses
display some non-universality at high scales; in fact, even when only
the third family displays this behaviour.  As non-universality in the
third family sector is relatively easy to accommodate vis a vis
flavour data, we adopt such a scenario to illustrate the viability of
such a situation.

It should be emphasized that, though we are illustrating the particle
spectrum under scrutiny in the context of a non-universal SUGRA, our real
stress is on the novel phenomenology which completely changes SUSY search
strategies. We all know that the most simple-minded SUGRA picture
(as well as many of its variants) is beset with a number of puzzles,
including issues related to FCNC. Of course, a generalization of SUGRA
can avoid most of such problems~\cite{leb2}. In general, however, one
is not sure that during SUSY searches at colliders one should adhere too
much to specific scenarios based on high-scale assumptions. Our ignorance
of, say, possible phenomena at intermediate scales further accentuates the
need of skepticism. In view of this, one feels that a consistent SUSY
scenario that leads to novel, unconventional experimental consequences is
worthy of investigation, irrespective of its high scale connection.
Something that can be tested at the early stage of the LHC is especially
interesting in this regard.

In the next section, we locate a few points in the parameter space
where a stop NLSP can coexist with a right-sneutrino LSP, on
introduction of non-universal scalar masses. The characteristic
signals at the LHC discussed here are (1a) $pp \longrightarrow 2$ {\rm
stop-tracks}, (1b) $pp \longrightarrow $ a single stop-track
accompanied by missing transverse energy and (2) $pp \longrightarrow $
1 or 2 stop-tracks, accompanied by multi-jets, missing transverse
energy and possibly some leptons.  We show that not
only are these signals separable from SM backgrounds but 
are also distinct from the signals
of a stau (or slepton) LSP. The discussions related to these signals,
together with the possibility of reconstructing gluino masses in this
scenario, are the contents of section 3. There we also comment on the
special consequences of the quasi-stable NLSP being colored (and
capable of hadronizing). We summarise and conclude in section 4.

\section{Right sneutrino LSP with a stop NLSP} 

\subsection{The scenario and some benchmark points}

With R-parity unbroken, the MSSM superpotential can be written as
\cite{moroi} 
\be W_{MSSM} = y_l L H_d E^c + y_d Q H_d D^c + y_u Q H_u
U^c + \mu H_d H_u 
\ee 
where $H_d$ and $H_u$  respectively are the Higgs doublets that give
mass to the down-type and up-type quarks. In the presence of the additional
neutrino superfields $N$, the superpotential can be extended by the
term
\be y_{\nu} L H_u N \ee 
where $y_{\nu}$ is given by $m_{\nu}=y_{\nu}\langle H_u
\rangle=y_{\nu}~ v~ \sin\beta$ with $v \, (\approx 246 \, {\rm GeV})$
being the electroweak symmetry breaking scale 
 and $\tan\beta=\langle H_u \rangle/\langle H_d \rangle$.
With the neutrino masses being atmost a few $eV$, we 
require $y_{\nu} \sin\beta \simeq 10^{-13} -10^{-12}$.

The general form of the sfermion mass matrix, neglecting 
inter-family mixing, can be written as
\be {M^2_{\tilde f}} = \left(
         \begin{array}{ll}
          m_{\tilde f_{LL}}^2     &  m_{\tilde f_{LR}}^2 \\
          m_{\tilde f_{LR}}^{2}   &  m_{\tilde f_{RR}}^2
         \end{array}
          \right) \ee   
where the diagonal elements are given by 
\be 
\barr{rcl}
\dis m_{{\tilde f_{LL}}}^2 & =  & \dis 
  m_{\tilde f_{L}}^2+ {m^2_Z} (T^{f}_{3L}-Q_f \sz) \cos2\beta + {m_f^2} 
\\
\dis m_{\tilde f_{RR}}^2 & = & \dis m_{\tilde f_{R}}^2+ 
Q_f {m^2_Z} \sz \cos2\beta + {m_f^2}
\earr
\ee
whereas the off-diagonal terms are 
\be
\barr{lcrcl}
\tilde u : & \qquad &
\dis m_{\tilde f_{LR}}^2 & =  & \dis 
    -m_f \, (A^f+\mu \cot\beta) = m_{\tilde f_{RL}}^2 
\\
\tilde d : & & 
\dis 
m_{\tilde f_{LR}}^2 & =  & \dis 
-m_f(A^f+\mu \tan\beta)= m_{\tilde f_{RL}}^2 \ .
\earr
\ee

	In a universal SUGRA scenario, all the low energy masses and
couplings can be expressed in terms of five free parameters defined at
the GUT scale, viz.  the universal scalar mass $m_0$, the universal
gaugino mass $m_{1/2}$, the universal trilinear soft SUSY-breaking
parameter $A_0$, the ratio of the vacuum expectation values of two
Higgses $\tan\beta$, and the sign of the Higgsino mass parameter $\mu$,
namely $sgn(\mu)$. The relevant parameters at the EW-scale are then
determined, via renormalization group evolution (RGE), from those
operative at the high scale of SUSY breaking in the hidden sector. Of
the resultant corrections to the (low-energy) squark and slepton
masses, the largest contributions accrue from the gauginos. The third
family masses also receive substantial corrections on account of the
Yukawa interactions and the mixing of left-and right-chiral states.

A right sneutrino LSP can be achieved in a part of the parameter space 
mostly favouring $m_0 < m_{1/2}$. With the one loop level RGE for 
$m_{\tilde{\nu}_R}$ given by
\be 
\frac{d m^2_{\rsnu}}{dt} = \frac{2}{16\pi^2}y^2_\nu~A^2_\nu \ ,
\ee
the smallness of the Yukawa interaction ($y_\nu \le 10^{-12}$), 
occasioned by our assumption of a conserved lepton number, serves to 
freeze the right-sneutrino mass at the high-scale value itself.
The lighter sneutrino mass eigenstate is given by 
\be 
\tilde{\nu}_1 = - \tilde{\nu}_L \sin\theta + 
\tilde{\nu}_R \cos\theta \simeq \rsnu \ ,
\ee
where the left-right mixing between the sneutrinos is given by
\be 
\tan 2\theta = \frac{2 \, y_\nu \ v \, \sin\beta \; |\mu \, \cot\beta -
A_\nu|}{m^2_{\tilde{\nu}_L}-m^2_{\rsnu}} \ .
\ee
Obviously, the state $\tilde{\nu}_1$ can become the LSP for a
sufficiently small value of $m_0$, and all the other particles in the
spectrum couple to it with a strength proportional to $y_{\nu}$. This
is so on account of the $\rsnu$ being a gauge singlet with the
consequence that its only interaction is via the Yukawa coupling. In
other words, any gauge coupling to $\tilde{\nu}_1$ depends on the
left-chiral component in it, which in turn again depends on $y_{\nu}$
(excepting for the pathological case where the two mass eigenstates
are degenerate to the level of $1$ in $10^{12}$).  Therefore, the
NLSP, irrespective of its identity, will decay into the
$\tilde{\nu}_1$ in an excruciatingly slow manner, making the former
appear stable in accelerator experiments.

Since our interest here is in a stop NLSP, we next identify points in
the SUGRA parameter space where this is possible.  In this, the
corresponding parameters should be allowed by the generic limits from
the direct search experiments (such as the Large Electron Positron (LEP) 
as well as the Fermilab Tevatron collider), and in particular should conform
to the specific bound on the mass of a quasi-stable stop. Furthermore,
they should also be consistent with other low-energy constraints such
as FCNC and with radiative breaking of the electroweak symmetry to
yield an acceptable vacuum.

The Tevatron Run IIb data for stop search, with $1 fb^{-1}$ integrated
luminosity, suggests that the lighter stop is constrained by $m_{{\tilde
t}_1}>220$ GeV \cite{tevuns}. In addition, in a recent simulation for
stable stop search at the Tevatron, as part of the Charged Matter
Stable Particles (CHAMP) analysis, it has been claimed that the
lighter stop should be above 250 GeV at $95\%$ confidence level
\cite{tevbo}. For our simulations, we have adopted a lower limit of 240
GeV for a quasi-stable stop.

To see if such a scenario can be realized within a universal SUGRA
setting, we performed a detailed study of the parameter space using
ISAJET 7.75 \cite{isa}. The simultaneous requirements of a stop NLSP
and a right sneutrino LSP yield only negative results. This is
because, in order to get a stop NLSP, one requires a large left-right
mixing which is driven by $A_t$ and $\cot\beta$. This is essentially 
to counter the large gluino contribution (in the RGE) from the 
top-gluino loop which is proportional 
to the gluino mass. The latter has to be large enough 
so that the mass of the lightest neutralino\footnote{Note that this constraint
would be relaxed if one were to admit nonuniversal gaugino masses at the 
high scale, thereby enlarging the parameter scale manifolds.} 
exceeds that of the
lighter stop ${\tilde t}_1$.  However, a large value of $A_t$ to
generate an effect of the above kind requires $A_0$ to be such as to
render some slepton (stau) tachyonic, or at any rate relegate it to
the level of the LSP.  Based on these considerations, a stop NLSP is
found very difficult to achieve in a universal SUGRA scenario.

\bt
\begin{center}
\begin{tabular}{|c|c|c|c|c|c|}                           \hline\hline
&{\bf Parameter}                      &{\bf BP1}    &{\bf BP2}   &{\bf BP3}    &{\bf BP4}\\\hline\hline
{\bf IN}&$m_0$, $m_{1/2}$, $A_0$               & 184,600,$-2400$&370,650,$-2600$&540,700,$-2500$&325,800,$-3000$\\
{\bf PUT}&$m_{\tilde{t}_L}$,$m_{\tilde{t}_R}$,&600,301 &700,400 &1000,200 &1000,260,750\\
(GeV)&$m_{\tilde{b}_R}=m_{\tilde{\tau}_L}=m_{\tilde{\tau}_R}$ &500 &750&750 &750\\\hline
&$|\mu|$                                     &1363 &1459 &1479 &1750 \\
&$\mel$,$\mml$                                 &461  &585  &743  &659  \\
&$\mer$,$\mmr$                                       &244  &415  &528  &336  \\
&$m_{\snu_{eL}}$,$m_{\snu_{\mu L}}$            &450  &576  &735  &648  \\
&$m_{\snu_{\tau_2}}$                         &581  &765  &1071 &865  \\  
{\bf O}&$m_{\snu_{eR}},m_{\snu_{\mu R}} $    &184  &370  &540  &325  \\
{\bf U}&$m_{\stau_1}$                        &316  &555  &871  &544  \\
{\bf T}&$m_{\stau_2}$                        &595  &775  &1077 &873  \\
{\bf p}&$m_{\chi^0_1}$                       &253  &276  &299  &342  \\
{\bf U}&$m_{\chi^0_2}$                       &485  &528  &571  &652  \\
{\bf T}&$m_{\chi^0_3}$                       &1359 &1455 &1478 &1756 \\  
&$m_{\chi^0_4}$                              &1361 &1457 &1481 &1748 \\
&$m_{\chi^{\pm}_1}$                          &488  &532  &574  &657  \\ 
&$m_{\chi^{\pm}_2}$                          &1363 &1459 &1483 &1750\\
&$m_{\tilde{g}}$                             &1367 &1477 &1594 &1790 \\  
&$m_{\tilde{u}_L},m_{\tilde{c}_L}$           &1260 &1391 &1530 &1653  \\
&$m_{\tilde{u}_R},m_{\tilde{c}_R}$           &1222 &1350 &1502 &1612  \\
&$m_{\tilde{d}_L},m_{\tilde{s}_L}$           &1263 &1394 &1532 &1655  \\
&$m_{\tilde{d}_R},m_{\tilde{s}_R}$           &1207 &1337 &1470 &1580  \\
&$m_{\tilde{t}_1}$                           &240  &273  &296  &330  \\ 
&$m_{\tilde{t}_2}$                           &1109 &1203 &1443 &1544  \\
&$m_{\tilde{b}_1}$                           &1075 &1174 &1423 &1534  \\
&$m_{\tilde{b}_2}$                           &1209 &1284 &1476 &1615 \\
&$m_{h^0}$                                   &116  &117  &121  &120  \\
&$m_{H^0}$                                   &1305 &1429 &1507 &1706  \\
&$m_{A^0}$                                   &1297 &1421 &1498 &1695 \\
&$m_{H^{\pm}}$                               &1308 &1432 &1510 &1708\\\hline
\hline
\end{tabular}
\caption{\small\it{Proposed Benchmark Points (BPs) for a stop NLSP in a
non-universal right-chiral sneutrino LSP SUGRA scenario. 
Non-universality in third generation sfermion masses has been assumed. Top 
mass is assumed to be $171.4$ {\cal GeV}. Values of all the mass parameters 
are in {\em GeV} units. Other SUSY parameters are: $\tan\beta=20$ and 
$sgn(\mu)=+$. Note that, $m_{\snu_{\tau_1}}$ can be fixed at any value below 
$m_{\tilde{t}_1}$.}}
\label{param_tbl}
\end{center}
\et

The spectrum of the type looked for, on the other hand, can still be
motivated in the SUGRA setting if some non-universality of scalar
masses at high scale is allowed. The type of non-universality sought
in our context is one where the third family sfermion masses are
different. Representative scenarios which can motivate such spectra
are those with additional U(1) symmetries (possibly anomalous) with
flavour-dependent D-terms \cite{nusf}, leading to arbitrary high-scale
soft masses for the stop, sbottom, stau and tau-sneutrino
states. However, rather than restricting ourselves to a particular model,
we perform a phenomenological analysis, and scan the parameter space
without any bias, to see if a stop NLSP can coexist with a (tau)
sneutrino LSP. Table \ref{param_tbl} contains four benchmark points
answering to such a description, on which our collider predictions are
based. The scan over the parameter space, using ISAJET 7.75, also
takes into account constraints such as those from LEP,
$b\longrightarrow s\gamma$ as well as the prospect of charge-and
colour-breaking vacuum and a vacuum unbounded from below. The magnitude of the 
Higgsino mass parameter $\mu$ has been fixed from
 electroweak symmetry breaking conditions, and the sign of $\mu$ (to which our 
results are not sensitive) has been taken as positive. The value of
the right-sneutrino mass does not affect the collider phenomenology in
any way unless it is heavier than the stop NLSP. We have thus kept it
as a free parameter, which can assume {\em any value compatible with
dark matter requirements}.

It should be mentioned here that a stop NLSP can also be achieved
with universal squark and slepton masses but different high-scale
$A$-parameters for the squark and slepton sectors. With $A_l
\ll A_q$, dangers such as tachyonic state modes can then be averted. Also, even
though we have ensured that processes such as, $b\longrightarrow s \gamma$
are within control with our parameter choice, a satisfactory suppression of
FCNC (including contributions to $B^0 - {\bar{B}}^0$ mixing) over a range
of parameters will require some model-dependent alignment mechanism for the
quark and squark mass matrices. Such a mechanism can keep the ``super-CKM''  
angles suitably small.

The proliferation of parameters in this scenario, which
is not surprising in a phenomenological study, makes it 
less illuminating than in a universal SUGRA to seek a pattern 
in the underlying high-scale physics. Nonetheless,
we notice the following general features in the
choices that give rise to the spectrum under study:

\begin{itemize}
\item
A large $|A_0|$ is required to generate a large left-right mixing in
the stop sector, so that a sufficiently large $m_{1/2}$ (required to
push up the lightest neutralino mass) can still be compatible with a
stop NLSP.

\item 
For a fixed (high scale) $m_{\tilde{t}_R}$, the allowed parameter
space becomes narrower as we increase $\tan\beta$. To push up
the down-sector sfermion-masses above the NLSP mass, we require 
large values of $m_0$ , while the need to place neutralinos above the 
NLSP implies a large $m_{1/2}$. A $\tan\beta$ in the range $5-35$ seems 
to be relatively more favorable for this purpose.
\end{itemize}

\begin{figure}[!h]
\begin{center}
\resizebox{70mm}{!}{\includegraphics{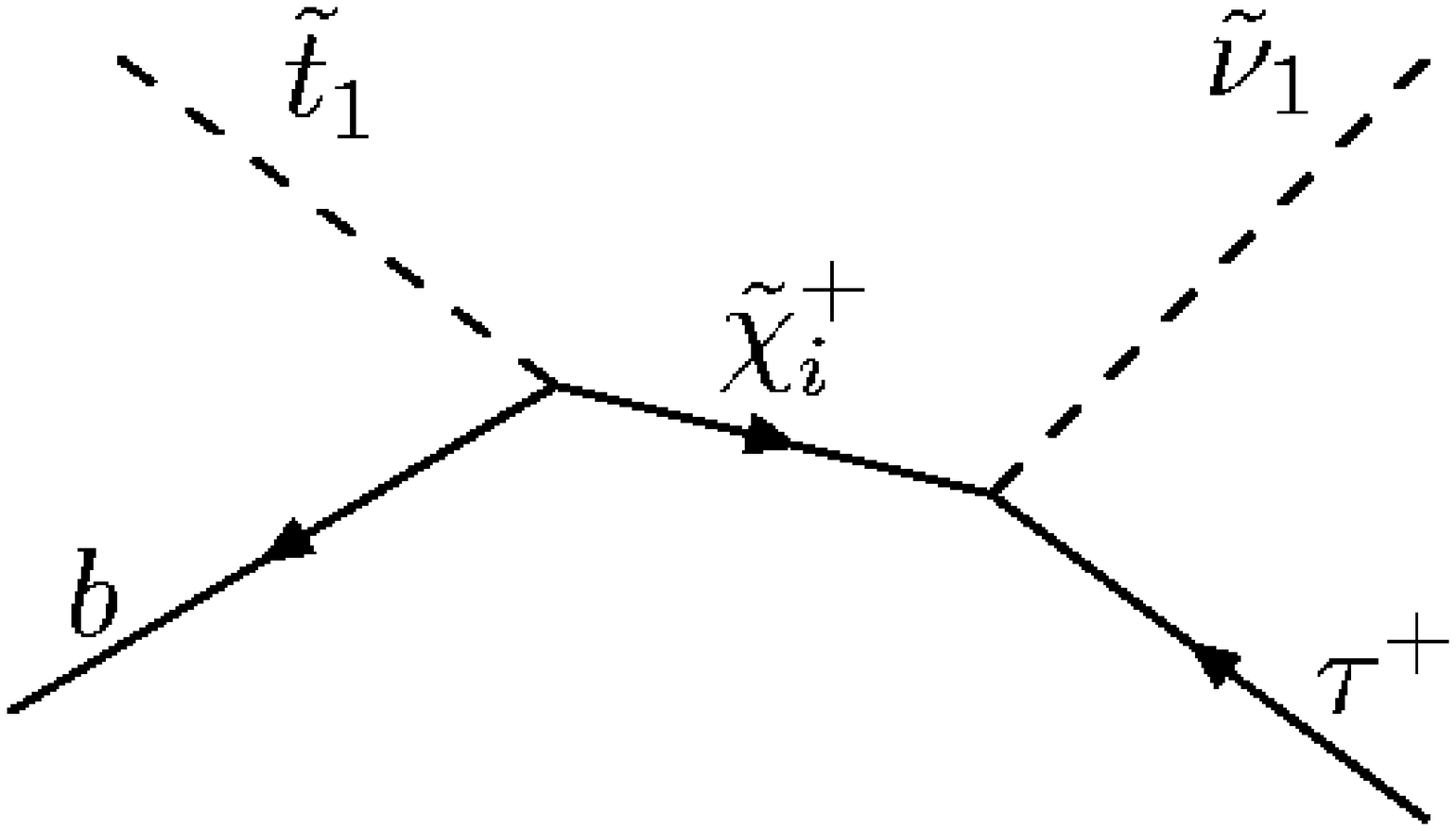}}
\hspace{.5cm} 
\resizebox{70mm}{!}{\includegraphics{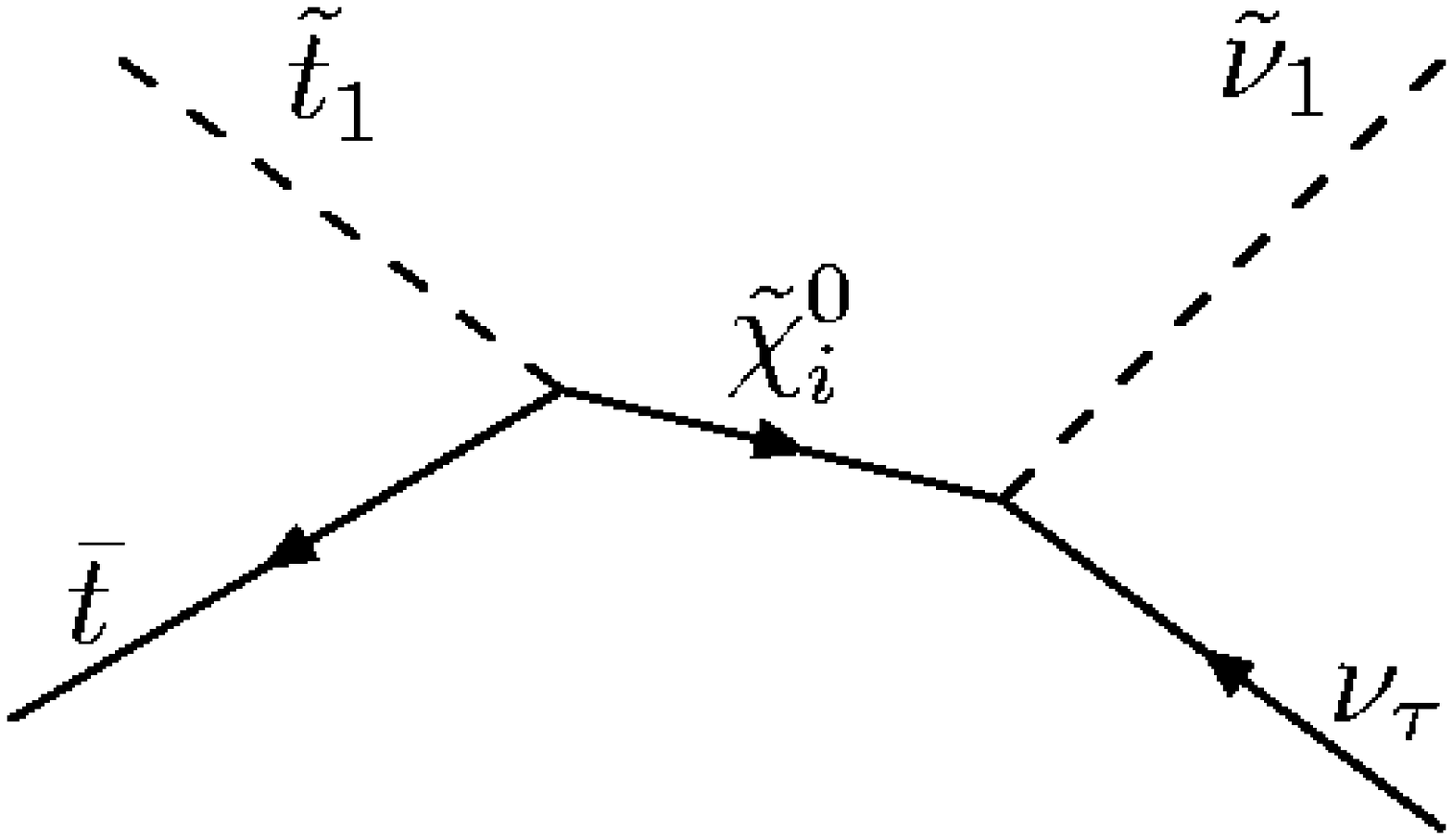}}
\caption{\small\it{Stop decay channels.}}
\label{dec}
\end{center}
\end{figure}

\subsection{Stop Life-time and stop-hadrons}
\label{sec:lifetime}

\begin{figure}[!h]
\begin{center}
\resizebox{70mm}{!}{\includegraphics{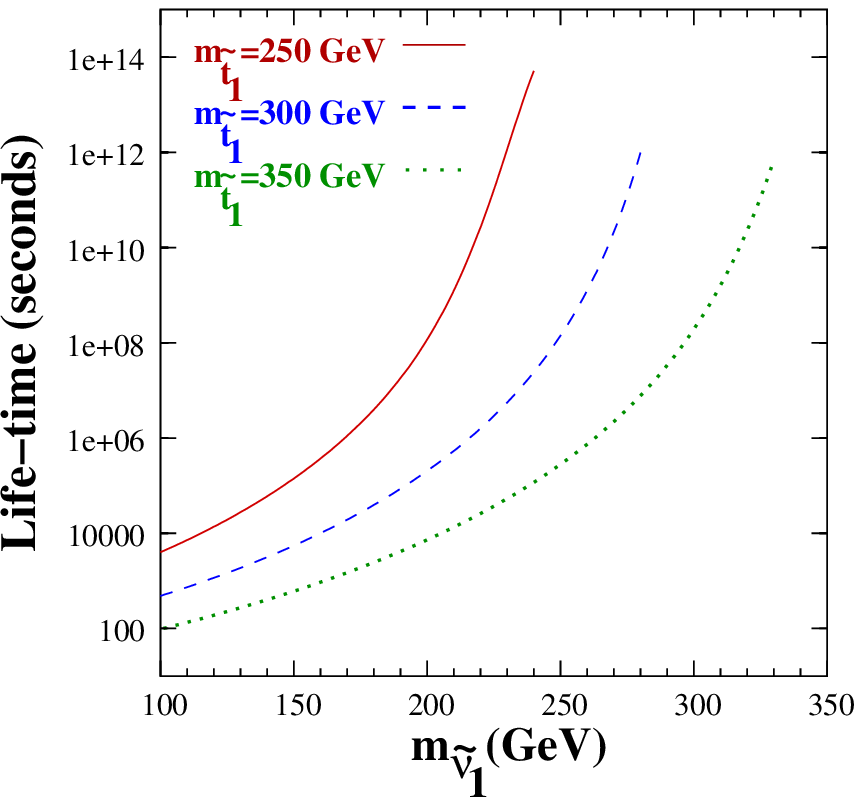}}
\hspace{.5cm} 
\resizebox{70mm}{!}{\includegraphics{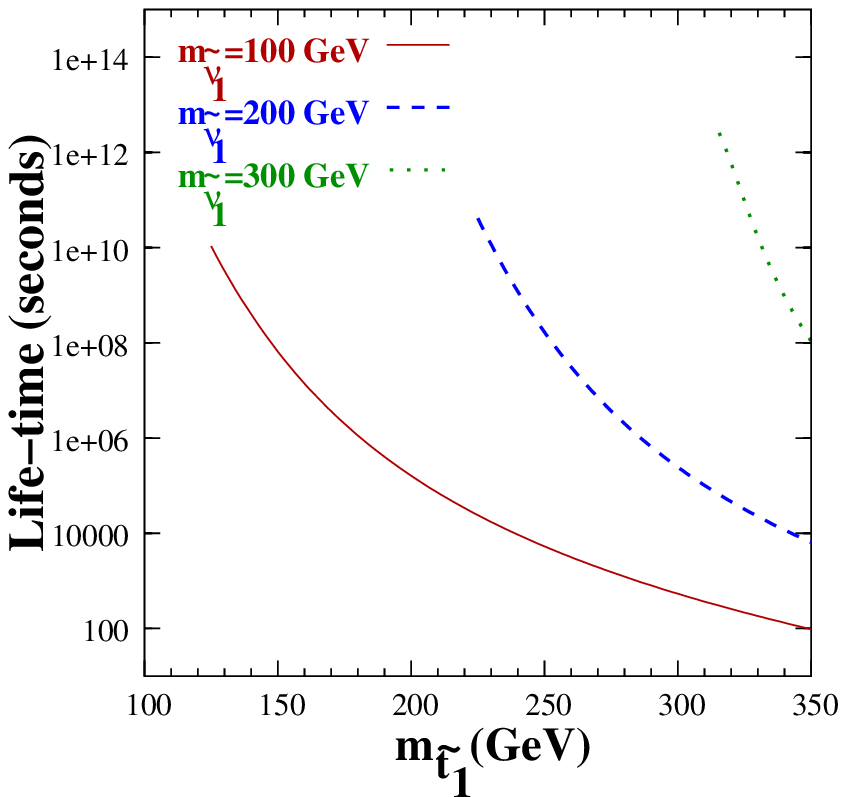}}
\caption{\small\em Rest frame life-time of the stop NLSP 
for {\em (a)} fixed NLSP mass, and, {\em (b)} fixed LSP mass.}
\label{lft}
\end{center}
\end{figure}

The major decay modes available to the $\stop_1$ NLSP are
$\stop_1\rightarrow b~ \snu_1~ \tau^+$ (via ${{\tilde{\chi}}_i}^\pm$)
and $\stop_1\rightarrow t~ \snu_1~ \nu$ (via 
${{\tilde{\chi}}_i^0}$). The corresponding Feynman diagrams are
presented in Fig.~\ref{dec}. The dependence of the decay rates on the
neutrino Yukawa couplings has already been discussed. In Fig.
~\ref{lft}, we present the decay lifetime 
for a wide range of LSP (NLSP) masses for a fixed NLSP (LSP) mass. The
lifetime rises with an increase in LSP mass whereas it understandably
decreases when the NLSP mass increases. The order of magnitude of the
$\stop_1$ lifetime shows rather unambiguously that, over a wide choice
of $\snu_1$ masses, the stop NLSP will decay way outside the
detector. A similar pattern in the lifetime plots of a stop NLSP with
a gravitino LSP has been reported earlier~\cite{yudi}. It is also to
be noted that though the NLSP is long-lived, its lifetime is always
smaller than the age of the universe, with, the present study
is safe from the viewpoint of charged dark matter. And, as long as the
lifetime is not too large ($\lsim 10^8$ s), one is safe from other
cosmological bounds such as those from big bang
nucleosynthesis~\cite{sandre}.

Even the minimal of lifetimes as in Fig.\ref{lft} imply that the stop
hadronizes before decaying. While the exact nature of fragmentation
characteristics would need to be worked out in detail, it is a very good
approximation to consider that half of the stops thus produced would
result in singly-charged stop-hadrons (say $\tilde t_1 \, \bar d$) while
the other half would result in neutrals ($\tilde t_1 \, \bar u$)
{\footnote{This can also happen 
in models with a stop LSP and a very small R-parity
violation, if one gives up on SUSY dark matter. For related work, see, for
example Ref.\cite{levey}.}}.  Although other hadrons, including doubly-charged
ones (such as $\tilde t_1 u u$) or excited states such as $\tilde t_1 \,
\bar d \, g$ are possible as well, fragmentation into them is suppressed
and deviations from the two-way splitting (with a 1:1 ratio) is expected
to be modified only to a very small extent. Furthermore, the very small 
mass difference between such hadrons implies that each of them
would be quasi-stable on the scale of the detector.

It has been argued~\cite{Arvanitaki:2005nq} though that the heavier of
these quasi-stable hadrons may decay strongly into the lighter ones
(and, in principle, cascade down) as long as they are kinematically
allowed to do so. For example,
if the charged hadron mass is larger than that of the 
corresponding neutral hadron by an amount exceeding the pion mass, such 
strong decays would 
cause the charged track(s) to disappear. On the other hand,
if neutral stop-hadrons are similarly heavier, then the charge tracks are 
produced and sustained, and the two tracks signal rates are enhanced over
what they have been found here.  While a definitive
statement can be made only on computing the spectrum of such
stop-hadrons, our experience with ordinary heavy-light quark bound
systems suggests that the mass difference between these two states
(which, presumably are the lightest of the stop-hadrons) would be well
below $m_\pi$, thus preventing a strong decay. The weak decay
lifetime, on the other hand, is much too long for it to be relevant
to collider studies.  While this argument would not hold for the
decays of, say $\tilde t \bar s$ or $\tilde t \bar u g$, the lower
fragmentation into these states renders such worries irrelevant at
the current level of sophistication.

Of more significance is the possibility that the stop-hadrons may
deposit some energy in the calorimeters through either quasi-elastic
or inelastic collisions~\cite{nusf}.  Various
claims~\cite{Arvanitaki:2005nq,Mohapatra:1999gg} and
counterclaims have been made in the literature in this regard.
In addition to the possibility of energy deposits by the
stop track in the hadron calorimeter, it is also possible
that the interaction with the calorimeter material will
convert charged stop-hadrons into neutral ones or vice versa \cite{shadrons}.
In this process, one may observe a charged track in the
inner tracking chamber, but no track in  the muon detector.
Alternatively, a neutral quasi-stable hadron with no record
in the inner tracker may get converted into a charged one
and display a track in the muon chamber, thus yielding signals
of a very novel type. A quantitative prediction of signals based on the above
observations will require (a) an elaborate detector simulation,
buttressed with data from initial run of the LHC, and (b) a reliable
model of hadronization of (quasi-)stable supersymmetric
particles. What we may conclude with a reasonable degree
of confidence is that the stop-hadron would deposit a small amount of
its energy in the hadronic calorimeter (differing from a quasi-stable stau
in this regard). Furthermore, the fraction deposited is generally
small enough to allow it to pierce through the muon chamber. 
Thus,
inspite of some quantitative uncertainties on this issue, one can
still predict a definite excess of signal over background, based on
rather simple assumptions. And, given the lack of a unambigious 
estimation of the conversion between stop-hadrons in matter, we 
deliberately choose to discount the novel signatures arising therefrom, 
limiting ourselves to the more conservative signals constructed 
solely with quasi-stable stop-hadrons that leave no trace when 
they are neutral and only a track when charged.

\section{Signatures of Stop NLSP at the LHC}

In the previous section, we observed that the stop NLSP will 
typically decay outside the detector. Thus, its collider
signatures will be in the form of charged tracks that show up in both
the inner tracker and the muon chamber. In general, the 
high velocities (note that stop production has a very large $P$-wave 
contribution) of these stable stops will make their identification
from time delays rather difficult. Although one can think in terms of
the thickness of the tracks and the small amount of energy deposit in
the hadron calorimeter, it is desirable to identify, instead, kinematic
characteristics that distinguish them. This is of paramount importance
since the most distinctive feature of SUSY in the minimal form,
namely $E_T{\!\!\!\!\!\!/\ }$, may be absent in a large fraction of the events 
in this scenario; yet the signals thereof may be striking for 
it is the tracks in the muon chambers that carry its imprints 
and truly characterize the scenario.

The most copious signal is the pair production of stable stops. This
yields a very large number of events of the type of Drell-Yan muon pair
production. However, as already mentioned, a stable stop will hadronize;  
we tentatively assume that its probability of forming a charged or neutral
hadron is $50\%$ each (see Sec.\ref{sec:lifetime}). Accordingly, one has
events with either one or two charged tracks events and these are
discussed in Sec.\ref{sec:tracks}.

\bt[!h]
\begin{center}
\begin{tabular}{|c|c|c|}\hline \hline
{\bf Signal}&{\bf Source}&{\bf Nomenclature}\\ \hline 
\hline
$2~charged-tracks$& distop-pair production & 1a\\
\hline		
$1~charged-track+ E_T{\!\!\!\!\!\!/\ }$& distop-pair production& 1b\\
\hline\hline
$2~charged-tracks+ 2~leptons + 2~jets + E_T{\!\!\!\!\!\!/\ }$& gluino-pair production & 2a\\
\hline
$2~charged-tracks+ 1~leptons + 4~jets + E_T{\!\!\!\!\!\!/\ }$&gluino-pair production & 2b\\
\hline 
$2~charged-tracks+ 0~leptons + 6~jets + E_T{\!\!\!\!\!\!/\ }$&gluino-pair production & 2c\\
\hline\hline
$1~charged-track+ 2~leptons + 2~jets + E_T{\!\!\!\!\!\!/\ }$&gluino-pair production & 3a \\
\hline
$1~charged-track+ 1~lepton + 4~jets + E_T{\!\!\!\!\!\!/\ }$&gluino-pair production &3b\\
\hline
$1~charged-track+ 0~leptons + 6~jets + E_T{\!\!\!\!\!\!/\ }$&gluino-pair production &3c\\
\hline\hline
\end{tabular}  
\caption {\small\it{A list of various signals with two and one charged 
track(s).}}
\label{channel_tbl}
\end{center}
\et
With the stop being considerably lighter than any of the other
strongly interacting sparticles, a gluino decays substantially into a
top and a stop. This leads to additional signals consisting of stable
stop tracks and a pair of top quarks produced in association (see
Table.\ref{channel_tbl}). Such signals have the advantage of
distinguishing stop tracks from those of stable staus. They can also,
in principle, enable one to reconstruct the gluino mass. We discuss
these signals in sections 3.2 and 3.3 respectively.

\subsection{Double and single-stop tracks}
\label{sec:tracks}

The main partonic processes responsible for this final state are 
$gg\rightarrow \stop_1 \stop_1^*$ and $q~\bar{q}\rightarrow \stop_1
\stop_1^*$. We use a CalcHEP-PYTHIA \cite{calc,pythia} interface for
our analysis, with CTEQ6L parton densities \cite{parton}. 
For the renormalization scale $\mu_R$ and factorization scale $\mu_F$, 
we use 
\be
\mu_R= 2 \, m_{\stop_1}=\mu_F 
\ee 
throughout the analysis. 
To obtain the next-to-leading order (NLO) results, we
multiply with the appropriate K-factor for the $\stop_1$ pair
production as computed in Refs. \cite{been1,been2}. The results presented
correspond to an integrated luminosity of $1 fb^{-1}$ at the LHC.

In order to get two charged tracks, each produced stop 
must hadronize to a charged hadron, thus reducing the rate by a 
factor of 4. For the two-track events, we use the following {\em basic 
cuts} at the outset:
\begin{itemize}
\item Each $\stop_1$ track should carry $p_T > 25$ GeV.
\item Both $\stop_1$'s should satisfy $|\eta|\leq 2.7$ to ensure 
that they lie within the coverage of the muon detector.
\item $\Delta R_{\stop_1\stop_1} \geq 0.2$ to ensure that the $\stop_1$'s
are well resolved in space.
\end{itemize}

\begin{figure}[bt]
\begin{center}
\resizebox{70mm}{!}{\includegraphics{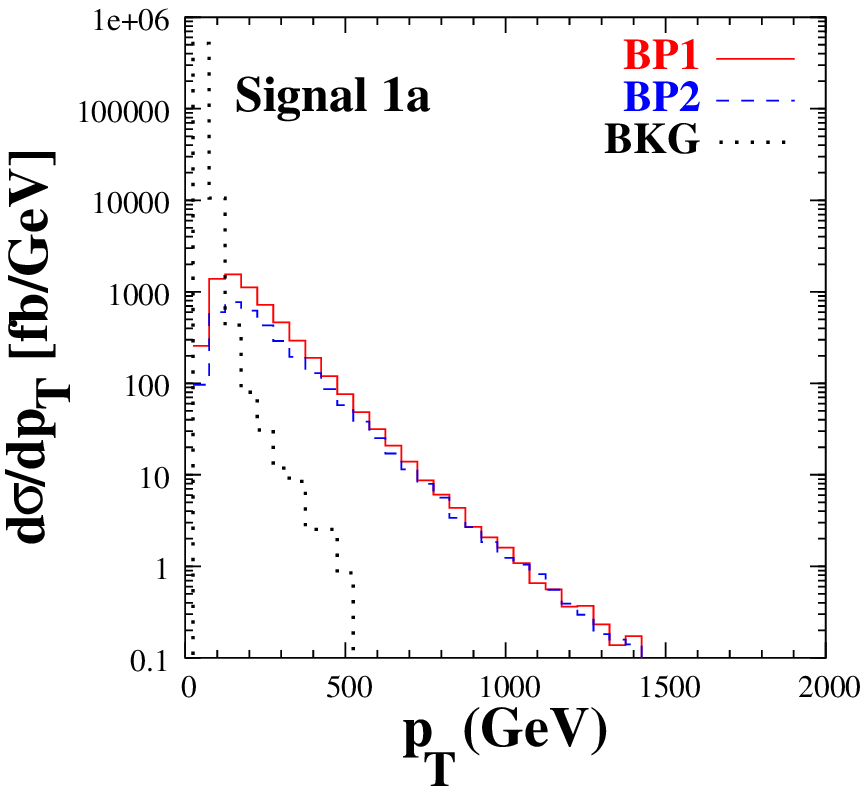}}
\resizebox{71mm}{!}{\includegraphics{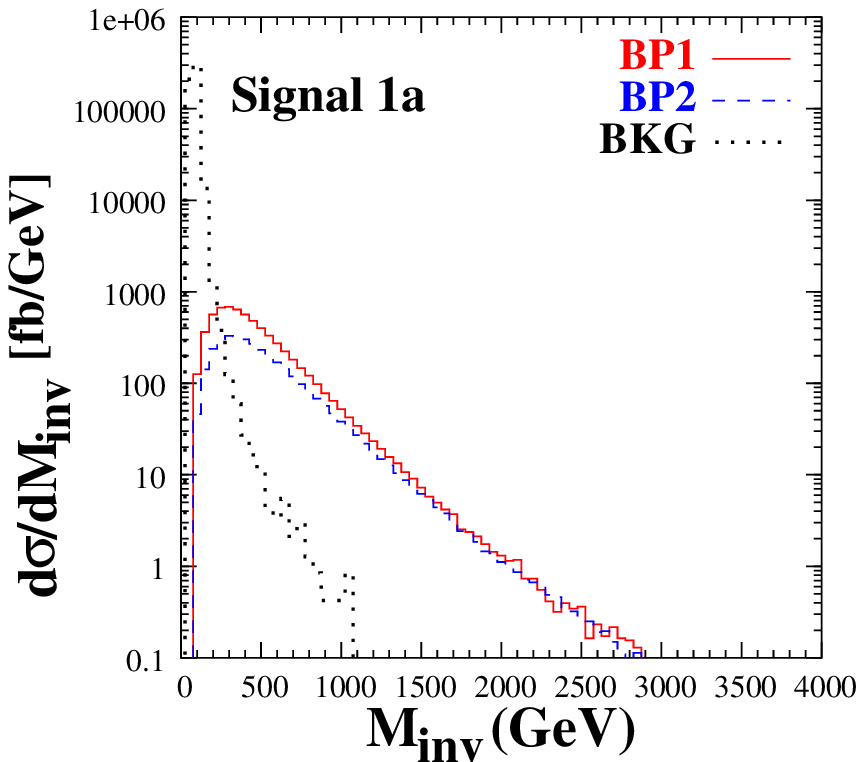}}

\caption{\small\it{$p_T$ (of the harder track) and track-pair invariant mass 
distributions with basic cuts for signal 1a.
Red (solid) and blue (dashed) histograms are for the signal in BP1 and BP2 
and black (dotted) histograms are for standard model background in both the plots.}}
\label{distop1a}
\end{center}
\end{figure}

The most important background \cite{bkg} to this signal comes from 
muon pairs produced in the Drell-Yan channel. 
The other source for the background is $WW$ pair 
production. We have also considered processes 
like $WZ$ and $ZZ$, giving rise to two detectable tracks in the muon 
chamber. There are still other sources such as triple gauge 
boson production, but the requirement of the invariant mass being 
sufficently above $m_Z$ will, in general, asphyxiate such events.

Assuming that the stop tracks are likely to be buried within
the copious backgrounds, we look for kinematic characteristics that 
can cause our predicted signal to stand out. With this in view, we 
show, in Figure \ref{distop1a}, the $p_T$ distributions
of the signal and the background. Also shown are the invariant mass
distributions of the pair of tracks, where the particles have
been assumed to be massless (so as to maximize the probability of
faking by Drell-Yan final states). Two out of the four benchmark
points have been chosen in each case, from which the general features 
are obvious.

It is clear from Figure \ref{distop1a} that most of the background 
muons are concentrated in the region of relatively low $p_T$. 
Therefore, an additional $p_T$ cut of 200 GeV has been imposed, which
suppresses the background significantly. 
In addition, a further cut on invariant mass on the pair of charged 
tracks, namely, $m_{\stop_1 \stop_1} >1100$ GeV completely removes the 
dimuon background. Note that the stop mass is unknown here, and the 
invariant mass is calculated from the track momentum, assuming that it is 
a massless particle. As the results demonstrate, this provides an effective 
event selection criteria for the signal. Thus a clean 
signature of the quasi-stable stop pair 
is obtained with an integrated luminosity $\int{\cal L}~ dt= 1 \, {\rm fb}^{-1}$
at the LHC\footnote{In fact, if the detectors are well understood, even 
a luminosity of 100 pb$^{-1}$ would be enough!}, 
as can be seen from Table ~\ref{distop1_tbl}.
As the same table shows, it is more efficient to use the combination
of the ($p_T ~+~ m_{\stop_1 \stop_1}$) cuts than just a higher $p_T$ cut
of 520 GeV, which is the softest one with which the background is 
completely gone.

\bt   
\begin{center}
\begin{tabular}{|c|c|c|c|c|c|c|}\hline \hline
{\bf Signal}&{\bf Cuts}&{\bf BP1}&{\bf BP2}&{\bf BP3}&{\bf BP4}&{\bf BKG}\\ \hline \hline
1a&$Basic$&6290&3390&2270&1320&$6.60 \times 10^3$\\\hline 
1a&$Basic+ p_{T} (\tilde t) \geq200$ GeV&1970&1290&970&645&104\\\hline
1a&$Basic+ p_{T} (\tilde t) \geq520$ GeV&119&99&87&71&0\\\hline
1a&$Basic + p_{T} (\tilde t) \geq200$ GeV &     &     &    &    &     \\
&$+ m_{{\tilde t}{\tilde t}}\geq 1100$ GeV&161&131&114&92&0\\ \hline\hline
1b&$Basic$&14000&7510&5030&2910&$5.75 \times 10^4$ \\\hline
1b&$Basic+ p_{T}({\rm track}) \geq200$ GeV&4060&2660&1990&1320&500\\\hline
1b&$Basic+ p_{T}({\rm track}) \geq200$ GeV  &&&&&\\
&$+ E_T{\!\!\!\!\!\!/\ } \geq400$ GeV&671&528&432&325&0\\ \hline\hline
\end{tabular}  
\caption {\small\it{The number of events---after various cuts---expected at 
the LHC
for signals 1a and 1b and for each of the benchmark points. The integrated
luminosity is assumed to be $1 fb^{-1}$. Also shown are the 
number of background events. Symbols have their usual meaning.}}
\label{distop1_tbl}
\end{center}
\et

\begin{figure}[bt]
\begin{center}
\resizebox{70mm}{!}{\includegraphics{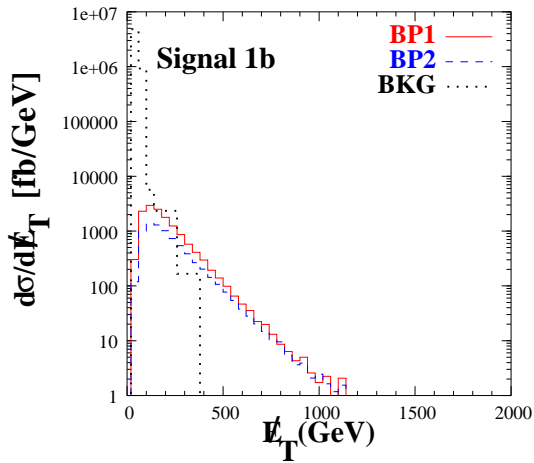}}
\caption{\small\it{$E_T{\!\!\!\!\!\!/\ }$ distributions with basic cuts for signal 1b.
Red (solid) and blue (dashed) histograms are for signal BP1 and BP2 
and black (dotted) histograms are for the standard model background.}}
\label{distop1b}
\end{center}     
\end{figure}

The single-track events, on the contrary, are associated with missing
$E_T$ assuming that the energy deposited by such superheavy neutral hadrons
in the hadron calorimeter is negligible\footnote{Similar conclusions are
drawn about the R-hadrons formed by long-lived gluinos in theories such as
split supersymmetry.}. The $p_T$ distribution of the track is the same as
in Figure \ref{distop1a}. Interestingly, a very similar distribution is
expected for $E_T{\!\!\!\!\!\!/\ }$ in this case (Figure \ref{distop1b}),
allowing deviations due to $p_T$-measurement only.

	Elimination of SM backgrounds (mostly from single $W$-production)
can be done by a procedure similar to the previous case. In this case, 
one can apply a $E_T{\!\!\!\!\!\!/\ }$ cut.
The results are shown in Table \ref{distop1_tbl}. One can see that there
are more signal events with zero background now. The reasons are (a) factor $2$
enhancement for the one charged track and one neutral track, (b) the
absence of any isolation requirements and (c) the $E_T{\!\!\!\!\!\!/\ }$
cut replacing the invariant mass cut for two charged tracks.

One may, however, like to ensure that these tracks are traced out
by a coloured particle such as a stop and not, for example, 
a pair of stable staus. With this in view, we have also considered
the production of stop tracks in cascades originating in gluino
pair production at the LHC, whose very nature distinguish the
tracks as those of squarks and not sleptons.

\subsection{Charged tracks from gluino production} 
 \label{sec:gluino-tracks}

In order to establish that these tracks are really due to stops (and not
stau's), we have studied signals 2 (with single charged track) and 3 (with
two charged tracks) listed in Table \ref{channel_tbl}. Such signals
can arise from gluino ($\tilde g$) pair-production, where both of the
gluinos decay into a (lighter) stop and a top, i.e. $\tilde g\rightarrow
\stop_1 t$. The different final state topologies arise due to leptonic or
hadronic decays of the $W$. For example, in case of signals 2(a) and 3(a)
both the $W$'s decay leptonically, for 2(b) and 3(b) one $W$ decays
leptonically whereas the other decays hadronically and, finally, 
in case of 2(c)
and 3(c) both the $W$'s decay hadronically. Thus from each top we will get
either one $b$-jet, one lepton and missing energy (due to neutrinos) or
one $b$-jet, and two other jets from the hadronic decay of a $W$. The decay
products of the two top quarks produced in association with the stops
establish the {\it bona fide} of the stop tracks. Although they are not
considered here, characteristic final states can be similarly chosen to
identify a sbottom NLSP. 

\bt   
\begin{center}
\begin{tabular}{|c|c|c|c|c|c|c|}\hline \hline
{\bf Signal}&{\bf Cuts}&{\bf BP1}&{\bf BP2}&{\bf BP3}&{\bf BP4}&{\bf BKG}\\ \hline \hline
2a&$Basic$&11&7&4&2&33\\\hline
2a&$Basic+\Sigma p_{T}\geq 800$GeV&11&7&4&2&0\\\hline\hline
2b&$Basic$&20&12&6&4&48\\ \hline
2b&$Basic+\Sigma p_{T}\geq 1200$ GeV&20&12&6&4&0\\\hline\hline
2c&$Basic$&35&25&14&9&98\\\hline
2c&$Basic+\Sigma p_{T}\geq 1500$ GeV&34&25&14&9&0\\\hline\hline
\end{tabular}  
\caption {\small\it{The number of events after various cuts for 
signals 2a-c at 
the LHC. The integrated
luminosity is assumed to be $300 fb^{-1}$. Symbols have their usual meaning. The 
$b$-tagging efficiency is not folded in.}}
\label{distop2_tbl}
\end{center}
\et

It should be remembered, however, that the gluino-induced signals
are not as abundant as in the previous case. The main reason for this 
is that we have assumed gaugino universality in our study. With
such an assumption, when the lightest neutralino is required to be heavier
than the lighter stop (whose mass in turn has to be at least about 
250 GeV from the CDF limits), the corresponding gluino mass is rather 
high leading to detectable but relatively small 
cross-sections\footnote{As mentioned earlier, non-universal gaugino 
masses could improve the situation dramatically}. 
The rates are further suppressed by branching ratios for specific
decays (with one or two leptons in final state) and the acceptance
cuts. Thus, in spite of the rather spectacular nature of the 
proposed signal, one has to struggle against statistics in general,
and higher luminosity is required. With this in view, we have made all
predictions for this class of signals with an integrated luminosity
of $300 fb^{-1}$.

The results of our analysis are presented in Tables~\ref{distop2_tbl} 
and~\ref{distop3_tbl}. The major sources for the backgrounds 
are tri-gauge bosons productions and $t{\bar t} l^+ l^{\prime -}$ 
(in case of two charged tracks), $t{\bar t} l \nu$ (in case 
of one charged track).

We work with the same basic cuts for these signals 
as mentioned in the previous subsection.
 In order that we are not inhibited by efficiency factors, we give up
$b$-tagging, which is not a serious disadvantage, in view of the
multiplicity of leptons (or muon-like tracks) in the final state. In
addition, we impose the following cuts:

\begin{itemize}
\item Each {\it jet} should have ${p_T}_j > 50$ GeV and $|\eta_j|\leq 2.7$.
\item $\Delta R_{\stop_1 j} \geq 0.4$, 
\item $\Delta R_{l j} \geq 0.4$,
\item Events must have missing energy $E_T >30$ GeV.
\end{itemize}

\begin{figure}[!h]
\begin{center}
\resizebox{70mm}{!}{\includegraphics{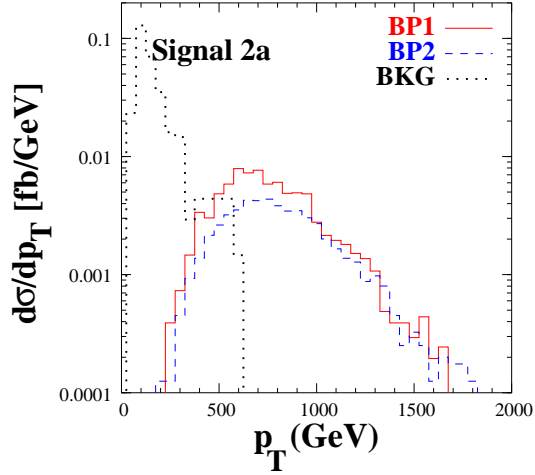}}
\caption{\small\it{$p_T$ (of the harder track) distributions with basic 
cuts for signal 2a.
Red (solid) and blue (dashed) histograms are for BP1 and BP2 
and black (dotted) histogram is 
for the standard model background. Symbols 
have their usual meaning.}}
\label{distop2a}   
\end{center} 
\end{figure}  

\begin{figure}[!h]
\begin{center}
\resizebox{55mm}{!}{\includegraphics{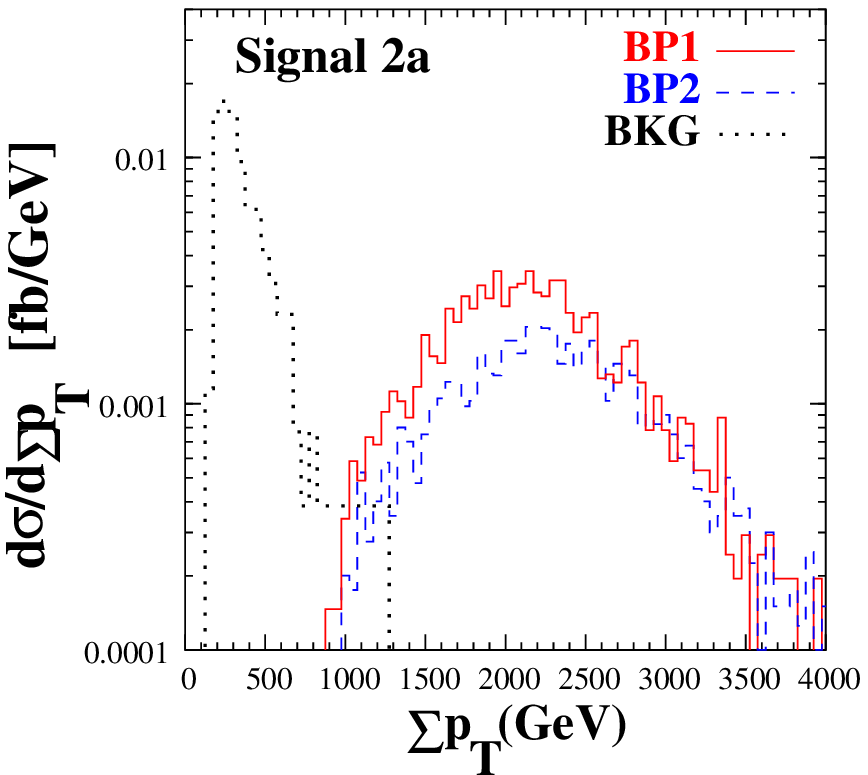}}
\resizebox{44mm}{!}{\includegraphics{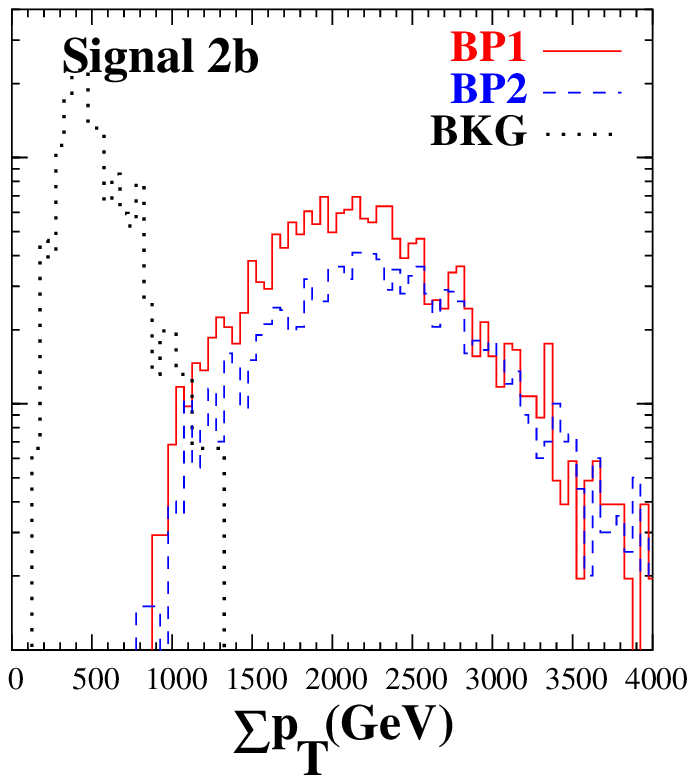}}
\resizebox{44mm}{!}{\includegraphics{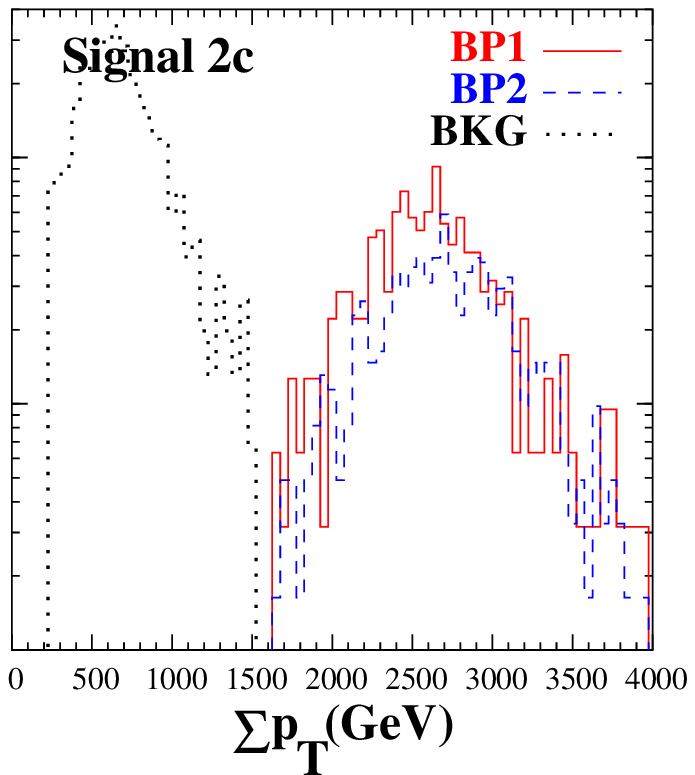}}
\caption{\small\it{Scalar summed $p_T$ distributions with basic cuts for signals 2a-c.
Red (solid) and blue (dashed) histograms are for BP1 and BP2 
and black (dotted) histogram is 
for standard model background in all the plots. Symbols have 
their usual meaning.}}
\label{distop2bc}   
\end{center} 
\end{figure}

\begin{figure}[!h]
\begin{center}
\resizebox{55mm}{!}{\includegraphics{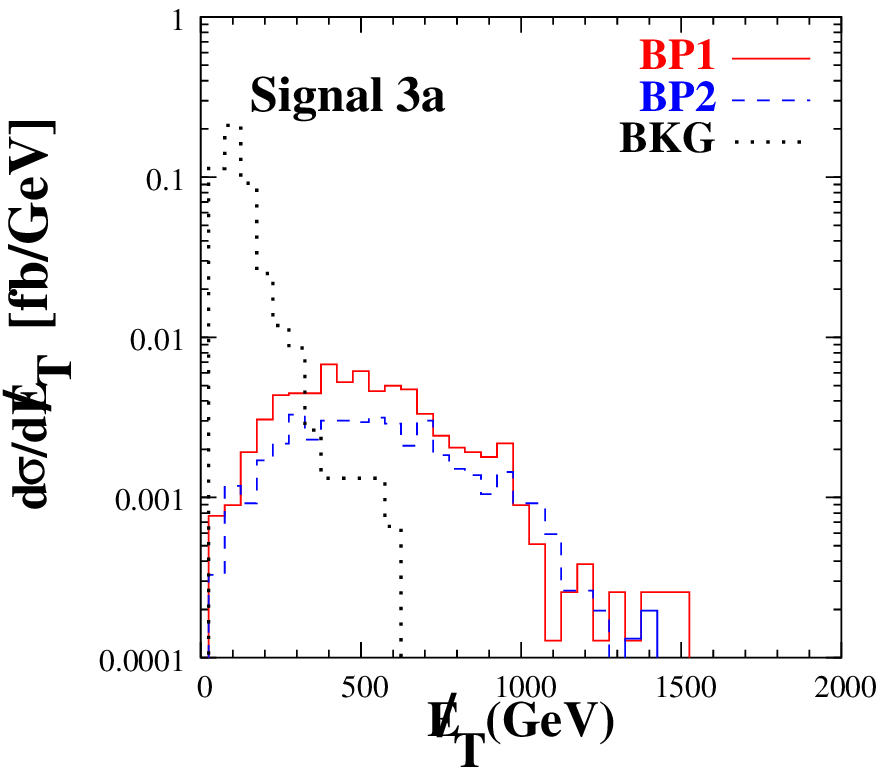}}
\resizebox{44mm}{!}{\includegraphics{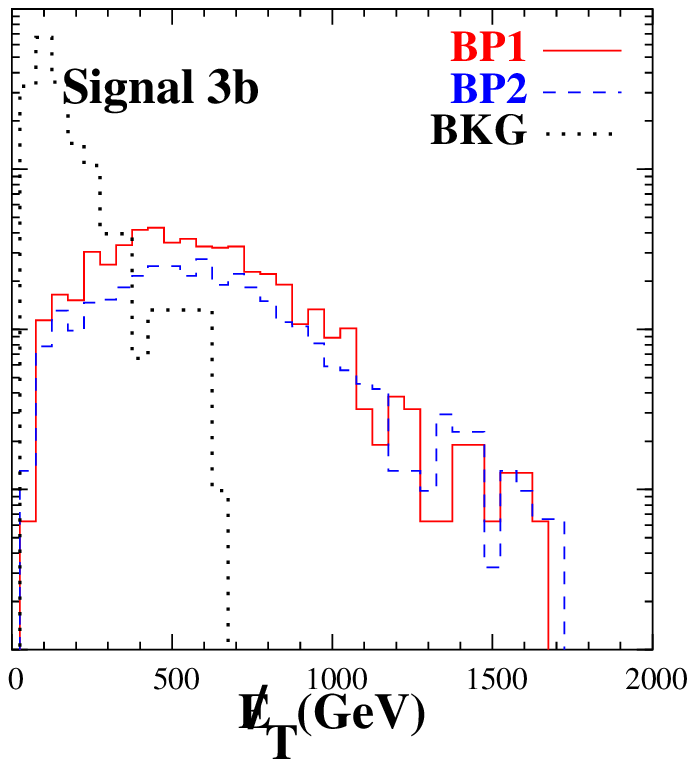}}
\resizebox{44mm}{!}{\includegraphics{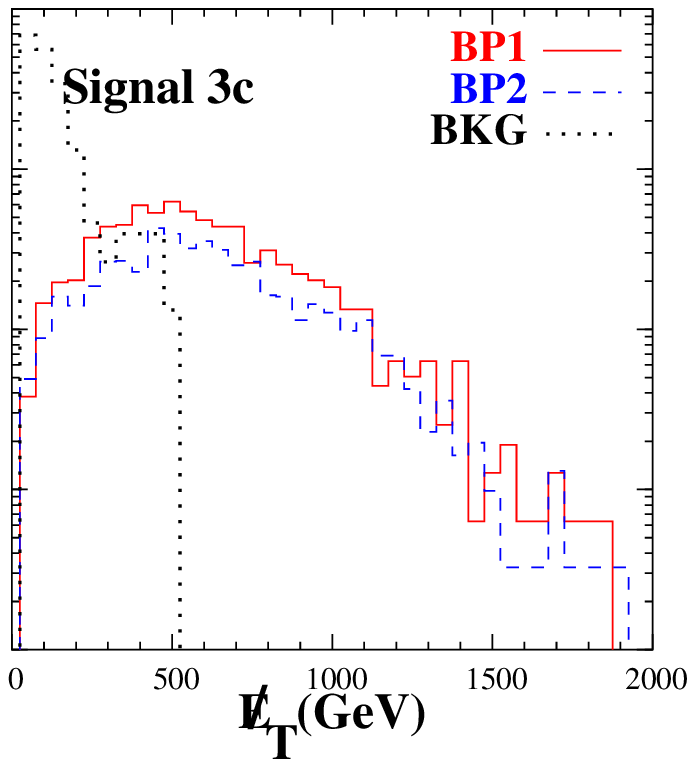}}
\caption{\small\it{$E_T{\!\!\!\!\!\!/\ }$ distributions with basic cuts for signals 3a-c.
Red (solid) and blue (dashed) histograms are for BP1 and BP2 and black (dotted) histogram is 
for the standard model background in all the plots. Symbols have their usual meaning.}}
\label{distop3abc}   
\end{center} 
\end{figure}

Figures \ref{distop2a} and \ref{distop2bc} contain plots of the
transverse momentum of the $\stop_1$
and the scalar sum of the the transverse momenta of all
visible particles for signals 2a-c (each with two charged tracks). Similarly,
missing energy distributions for signals (3a-c) with one charged track
are also shown in Figures \ref{distop3abc}. The corresponding plots for
the background are also shown.
  
It is found that a cut on the scalar sum of transverse momenta of visible
particles, namely, $\Sigma p_T >800, 1200$ and $1500$ GeV removes the
background completely in case of signal 2a, 2b and 2c respectively. The
corresponding requirement in case of each of signals 
3a-c is a missing energy cut of
$E_T{\!\!\!\!\!\!/\ } > 600$ GeV. The efficiency of
these cuts for all the four benchmark points is demonstrated in Table
\ref{distop2_tbl} for signals 2a-c and in Table \ref{distop3_tbl} for signals
3a-c. It should be noted that the low event rate due to branching fraction
suppression implies that such signals requires {\it $\int{\cal L}~ dt= 300
fb^{-1}$}. Clearly, $b$-tagging will destroy the detectability of BP4, but
not for the other benchmark points 
 in case of signals 2a-c and $3a$.

\bt[!h]
\begin{center}
\begin{tabular}{|c|c|c|c|c|c|c|}\hline \hline
{\bf Signal}&{\bf Cuts}&{\bf BP1}&{\bf BP2}&{\bf BP3}&{\bf BP4}&{\bf BKG}\\ 
\hline \hline
3a&$Basic$&21&13&7&4&140\\ \hline
3a&$Basic+ E_T{\!\!\!\!\!\!/\ } \geq600$ GeV&8&6&4&2&0\\ \hline\hline
3b&$Basic$&155&102&55&32&519\\\hline
3b&$Basic+ E_T{\!\!\!\!\!\!/\ } \geq 600$ GeV&62&42&28&18&0\\ \hline\hline
3c&$Basic$&236&148&79&49&558\\\hline
3c&$Basic+ E_T{\!\!\!\!\!\!/\ } \geq 600$ GeV&94&64&45&31&0\\ \hline\hline
\end{tabular}  
\caption {\small\it{The number of events after various cuts for 
signals 
3a-c at the LHC. The integrated luminosity is assumed to be $300 fb^{-1}$. Symbols have 
their usual meaning.}}
\label{distop3_tbl}
\end{center}
\et

\subsection{Gluino Mass reconstruction}

Since the quasi-stable stop is visible in this scenario,
a variant of the signal discussed in the previous subsection
can be used for the direct reconstruction of the gluino mass.
Note that this is very difficult to achieve in the minimal SUGRA 
scenarios on account of the fact that each supersymetric production event 
results in a pair of (invisible) LSPs being produced.  

The dominant decay mode involves both tops going hadronically
resulting in as many as six jets along with stop-track(s) and/or
missing transverse energy. Although it is possible, in principle, to
use such a final state for this purpose, it is normally beset with
problems and the attendant loss in accuracy. In our study, therefore, we
shall omit this channel altogether and concentrate on subdominant
modes even at the cost of signal strength. In other words, we only
consider the case where, of the two top quarks produced from a gluino
pair, one decays hadronically and the other leptonically.

\subsubsection{Two stop-tracks}

If both stops hadronize into charged tracks, the signal becomes 
\[
pp \longrightarrow 2 \, \hbox{\rm stop-tracks} + 1 {\rm \ lepton }
+ 2 \, b + 2 \, {\rm jets} + E_T{\!\!\!\!\!\!/\ } \ .
\]
The successful removal of 
backgrounds due to $t \bar{t} l \nu_l$, and also the suppression of a rather
sizable combinatorial background, prompts us to advocate
$b$-tagging in this case.

To be able to reconstruct the gluino mass, we need to assume that the
entire missing transverse energy in such events accrues from a single
invisible particle in the final state, namely, the neutrino. Using
energy and momentum balance in the transverse plane, and the fact that
the neutrino arises from a $W$ (of known mass), one can then
reconstruct the longitudinal component of the neutrino momentum (and,
thus, of the $W$) upto a two-fold ambiguity. The second $W$ is
completely reconstructed through hadronic decays. This, then, allows
us to reconstruct both the tops without any ambiguity (on insisting that 
the two tops thus reconstructed should have the same mass upto 
measurement and resolution uncertainties).

Next, we face a further combinatorial ambiguity, namely that arising
from the correct identification of the top-stop pairings. Note,
though, that the charge of each stop track is measurable and that a
$\tilde t \, (\tilde t^*)$ would, in general, be associated with a
positively (negatively) charged track. Thus, if the lepton were
positively (negatively) charged, the corresponding top (anti-top)
should be paired with negatively (positively) charged track. However,
since the gluino is a Majorana fermion, both stop tracks can be of the
same charge in $50 \%$ cases. This uncertainty as well as a two-fold
ambiguity due to the neutrino can be removed by demanding that the two
gluino masses, thus reconstructed, should not differ by more than $50$
GeV. In this manner, one can throw out the wrong combinations and
reconstruct the gluino peak.

We may now use the same basic cuts as those suggested in the previous
subsection. To make the reconstruction as clean as possible, we
require $\Sigma p_T >1200$ GeV. Table \ref{distop3_tbl} shows that
the backgrounds can still be eliminated by this method, although the
number of events is less than in the previous case, due to b-tagging
(with an assumed efficiency of 60\% \cite{btag}).

\begin{figure}[!h]
\begin{center}
\resizebox{82mm}{!}{\includegraphics{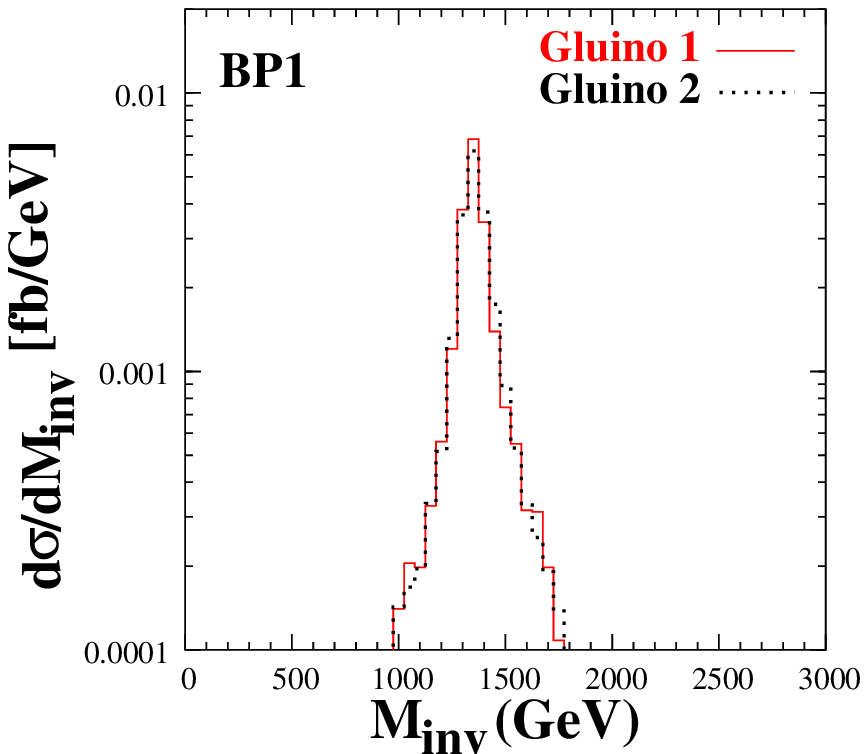}}
\resizebox{66mm}{!}{\includegraphics{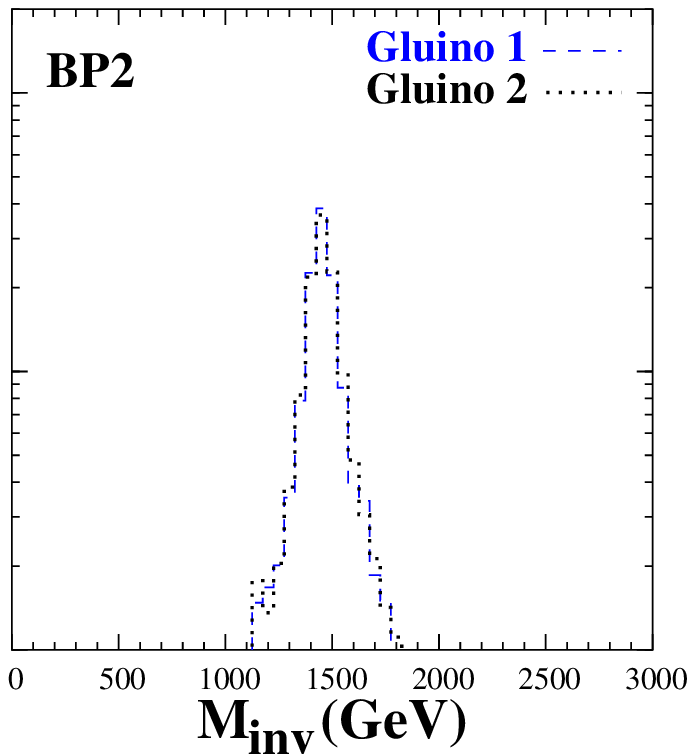}}
\caption{\small\it{Reconstructed invariant mass peaks for two gluinos for BP1 and 
BP2 in the signal $pp \longrightarrow 2 charged-track +
1 lepton + 2 b + 2 jets + E_T{\!\!\!\!\!\!/\ }$.}}
\label{glma_1}
\end{center}
\end{figure}

The results of this procedure for two of our four benchmark points 
are presented in Figure \ref{glma_1}, which show that the gluino mass 
can be reconstructed with about $10\%$ uncertainty. The event rates 
corresponding to the two remaining benchmark points are even lower 
(as seen for from Table \ref{distop2_tbl}.

On the whole, though the method described above works in principle 
(and barring the 6-jet final state, is perhaps the best option) 
for the said channel, it suffers from the problem of poor statistics. 
To enhance the number of events, we now explore the other 
channel, namely, where one of the two stops from gluino decay is
invisible, and investigate its usefulness in gluino mass
reconstruction.

\subsubsection{One stop-track}

With one stop going to a charged supersymmetric-hadron and the other into 
a similar neutral hadron, the number of events in this channel would be 
at least twice as many as in Sec.\ref{sec:gluino-tracks}. 
The signal now is 
\[
pp \longrightarrow 1 \, {\rm stop-track} + 1 {\rm \ lepton }
+ 2 \, b + 2 \, {\rm jets} + E_T{\!\!\!\!\!\!/\ } \ ,
\]
where the missing transverse energy now has two irreducible sources,
namely the neutral s-hadron and the neutrino from the top-decay. Once again,
$b$-tagging is needed. 

\begin{figure}[!h]
\begin{center}
\resizebox{82mm}{!}{\includegraphics{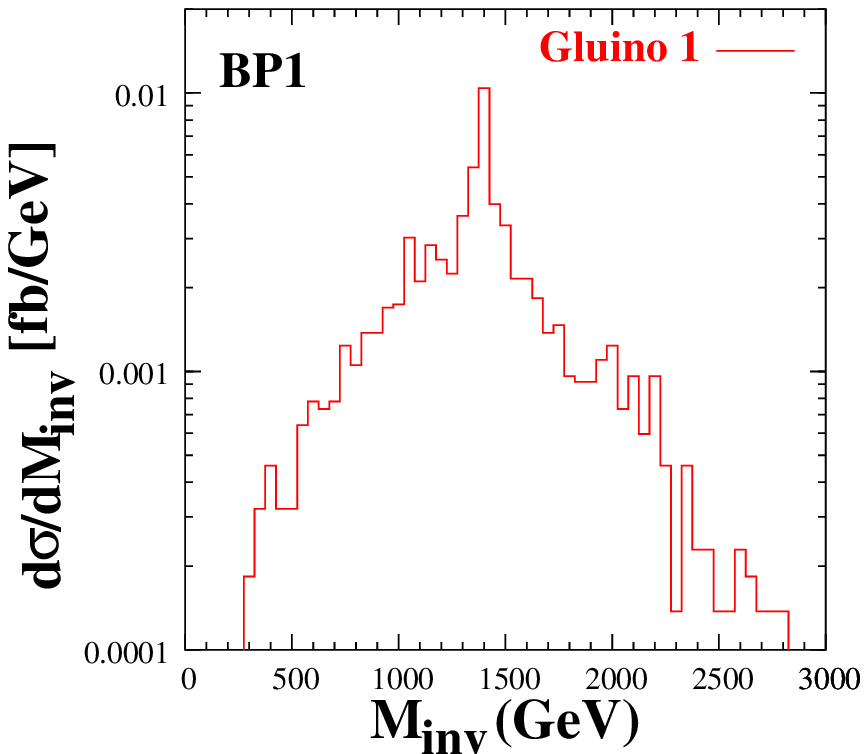}}
\resizebox{66mm}{!}{\includegraphics{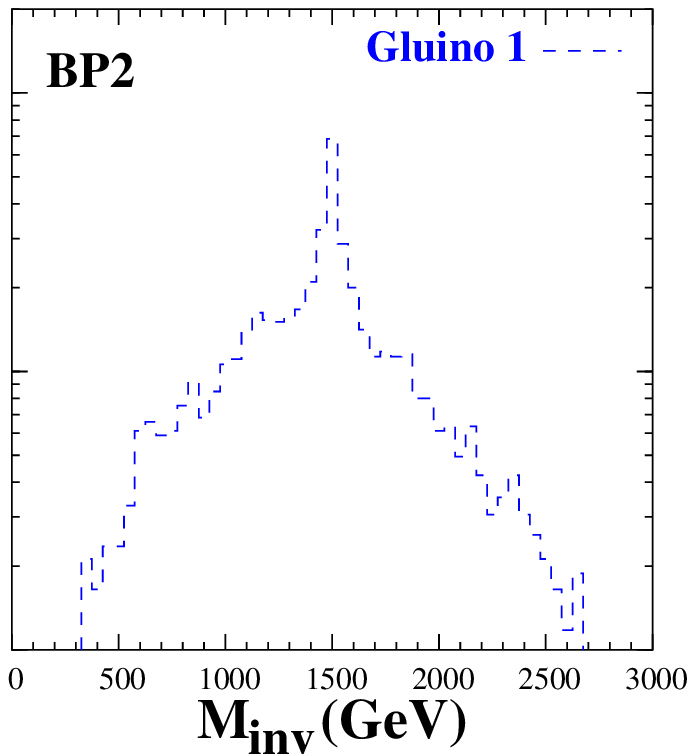}}
\caption{\small\it{Reconstructed invariant mass for one gluino for BP1 and BP2 in the 
signal $pp \longrightarrow 1 charged-track + 1 lepton + 2 b + 2 jets + E_T{\!\!\!\!\!\!/\ }$.}}
\label{glma_2}  
\end{center}  
\end{figure}

The reconstruction of the hadronically decaying top proceeds as in the
previous subsection. For obvious reasons, the reconstruction of the
leptonically decaying $W$, and hence the parent (second) top, cannot
be done now. The key step, then, is to decide whether the
reconstructed top came from the same gluino as the visible stop
track. In the absence of such a decision algorithm, the naive
procedure would be to forcibly associate the two and consider the
resultant invariant mass. The `correct' cases (where the $b + 2 \,
{\rm jets}$ system yields the top mass), then, would be expected to
lead to a concentration of events near the true mass (modulo
resolution effects) while the wrong identifications would lead to a
scattered distribution. The resultant is displayed in
Fig.\ref{glma_2}.

Were gluinos not Majorana particles~\cite{Kribs:2007ac} , the situation
could have been easily improved by the charge identification method
suggested above. Note that the sign of the charge of the visible
stop-track is easily measurable and corresponds almost uniquely to the
charge of the stop. Similarly, the sign of the lepton uniquely determines
the sign of the top decaying leptonically.  Thus, for the stop-track to
have arisen from the same parent Dirac-gluino as the reconstructed top,
the sign of its charge would have to be the same as that of the lepton.
Unfortunately, though, the Majorana nature of the gluino precludes such an
association, and the opposite charge combination (for stop and lepton) is
as likely to occur as the same-sign one.

We may now attempt to combine the significance of both methods to get the
final resolution on the gluino mass. It should be noted that the stop
track has been assigned zero mass in the reconstruction algorithm, in
spite of which the peaks are recovered quite accurately, modulo the
statistics in each case.

\section{Summary and conclusion}

We have investigated the signals of a stop NLSP in a scenario where the
LSP is a right sneutrino, with the stop decay into the LSP taking place
outside the detector. After convincing ourselves that such a scenario can
arise in SUGRA with non-universality in third family sfermion masses, we
have identified a few benchmark points, allowed by all the electroweak and
dark matter constraints, where the long-lived stop NLSP can be visible in
the form of charged tracks in the muon chamber. We have analyzed different
signatures of such tracks at the LHC, suggesting acceptance cuts with
which one can remove standard model backgrounds effectively. Final states
with {\it two charged tracks} (where a pair of stops both hadronize into
charged hadrons) and those with {\it one visible track} have been studied
in this spirit. It is found that one can have enough signal events with no
background, with an integrated luminosity of 1 $fb^{-1}$ or even less, so
that such a new physics signal cannot be missed.

In fact, even for the initial run of the LHC at 10 TeV, there is hope for
having the first hints of such a scenario if it exists. For BP1, for
example, our estimate predicts about 5 events for signal 1a, and for BP2,
4 events, with $\sqrt{s} = 10$ TeV, an integrated luminosity of 100
$pb^{-1}$ and the same cuts as in reported earlier. For signal 1b, about
twice as many events in each case can be expected. Since a reduction in
the centre-of-mass energy means the tracks slightly softer, the background
is absent in these cases even with the same cuts. Thus signals 1a and 1b
are predicted at the discovery level for the 10 TeV run, if
$\int{\cal{L}}dt$ = 100 $pb^{-1}$ is attained.

Moreover, the stop track can be distinguished from a slepton or stau track
(or that of a long-lived squark of the first two families) through gluino
decay into a top and a stop, and stable tracks produced in association
with a pair of top quarks. However, for the region of parameter space that
is phenomenologically consistent, the event rate is smaller than that in
the previous case, and one may require an integrated luminosity of 300
$fb^{-1}$. It is also possible to use the long-lived stops to reconstruct
the gluino mass, so long as it is within about 1.5 TeV.

It should also be borne in mind that the rather poor statistics expected 
in the channel used for gluino reconstruction is due to the fact that
we are adhering to a scenario with gaugino mass unification. 
The unification conditions requires the gluino to be rather heavy, and
therefore the production rates correspondingly suppressed, since the
lightest neutralino (to whose mass the gluino is related by the
unification condition) is to be higher than the lighter stop. 
However, such a restriction does not apply to a situation where gaugino
universality is either absent, or the Grand Unification group is broken
by some non-trivial representation~\cite{BM}. A relatively
lighter gluino in any of these `non-universal' cases is bound to 
push up the event rates for gluino pair production considerably,
and one has much better hopes of their reconstruction if a sneutrino
LSP scenario prevails. In fact, this is one reason why we have discussed
our suggested reconstruction techniques so elaborately.

Further studies related to spin measurement of such a stop 
NLSP can be worthwhile, thus providing clues on whether the tracks 
can be faked by some long-lived fermion. However, such a
study is beyond the scope of the present work.


\vspace*{0.4cm} \noindent {\bf Acknowledgments:} 
We thank AseshKrishna Datta for useful discussions. 
The work of SKG and BM was partially supported by funding available 
from the Department of Atomic Energy, Government of India, for the Regional 
Centre for Accelarator-based Particle Physics, Harish-Chandra 
Research Institute. DC acknowledges support from the Department of Science 
and Technology, India under project number SR/S2/RFHEP-05/2006. 
Computational work for this study was partially carried out at the cluster 
computing  facility in the Harish-Chandra Research
Institute ({\tt http:/$\!$/cluster.mri.ernet.in}). 
 
 \vskip 5pt


\begin{thebibliography}{999}

\bibitem{susyrev}
  For reviews see, for example, 
  H.~P.~Nilles,
  Phys.\ Rept.\  {\bf 110}, 1 (1984);
  H.~E.~Haber and G.~L.~Kane,
  Phys.\ Rept.\  {\bf 117}, 75 (1985);
  M.~Drees,
  arXiv:hep-ph/9611409;
  S.~P.~Martin,
  arXiv:hep-ph/9709356, and references therein;
  D.~J.~H.~Chung, L.~L.~Everett, G.~L.~Kane, S.~F.~King, J.~D.~Lykken
  and L.~T.~Wang,
  Phys.\ Rept.\  {\bf 407}, 1 (2005)
  [arXiv:hep-ph/0312378].

\bibitem{revnu}
For a review, see for example,
  R.~N.~Mohapatra {\it et al.},
  arXiv:hep-ph/0510213 and references therein.

\bibitem{rneut}
  T.~Hebbeker,
  Phys.\ Lett.\ B {\bf 470}, 259 (1999)
  [arXiv:hep-ph/9910326];
  N.~Arkani-Hamed, L.~J.~Hall, H.~Murayama, D.~R.~Smith and N.~Weiner,
  Phys.\ Rev.\ D {\bf 64}, 115011 (2001)
  [arXiv:hep-ph/0006312];
  A.~T.~Alan and S.~Sultansoy,
  J.\ Phys.\ G {\bf 30}, 937 (2004)
  [arXiv:hep-ph/0307143];
  D.~Hooper, J.~March-Russell and S.~M.~West,
  Phys.\ Lett.\ B {\bf 605}, 228 (2005)
  [arXiv:hep-ph/0410114];
  T.~Asaka, K.~Ishiwata and T.~Moroi,
  Phys.\ Rev.\ D {\bf 73}, 051301 (2006)
  [arXiv:hep-ph/0512118].

\bibitem{moroi}
  J.~McDonald,
  JCAP {\bf 0701}, 001 (2007)
  [arXiv:hep-ph/0609126];
  T.~Asaka, K.~Ishiwata and T.~Moroi,
  arXiv:hep-ph/0612211.

\bibitem{c8thermal}
  V.~Page,
  JHEP {\bf 0704}, 021 (2007)
  [arXiv:hep-ph/0701266].

\bibitem{drees}
  For an earlier discussion on such situations, see, for example,
  M.~Drees and X.~Tata,
  Phys.\ Lett.\  B {\bf 252}, 695 (1990).

\bibitem{stau}
  S.~K.~Gupta, B.~Mukhopadhyaya and S.~K.~Rai,
  Phys.\ Rev.\  D {\bf 75}, 075007 (2007)
  [arXiv:hep-ph/0701063].

\bibitem{gopal}
  C.~L.~Chou and M.~E.~Peskin,
  Phys.\ Rev.\ D {\bf 61}, 055004 (2000)
  [arXiv:hep-ph/9909536];
  A.~de Gouvea, S.~Gopalakrishna and W.~Porod,
  JHEP {\bf 0611}, 050 (2006)
  [arXiv:hep-ph/0606296].
  
\bibitem{yudi}
  J.~L.~Diaz-Cruz, J.~R.~Ellis, K.~A.~Olive and Y.~Santoso,
  JHEP {\bf 0705}, 003 (2007)
  [arXiv:hep-ph/0701229].

\bibitem{bkg}
  B.~Mele, P.~Nason and G.~Ridolfi,
  Nucl.\ Phys.\  B {\bf 357}, 409 (1991);
  R.~Hamberg, W.~L.~van Neerven and T.~Matsuura,
  Nucl.\ Phys.\  B {\bf 359}, 343 (1991)
  [Erratum-ibid.\  B {\bf 644}, 403 (2002)];
  T.~Binoth, M.~Ciccolini, N.~Kauer and M.~Kramer,
  JHEP {\bf 0612}, 046 (2006)
  [arXiv:hep-ph/0611170].

\bibitem{leb2}
  P.~H.~Chankowski, O.~Lebedev and S.~Pokorski,
  Nucl.\ Phys.\  B {\bf 717}, 190 (2005)
  [arXiv:hep-ph/0502076].


\bibitem{tevuns}
T.~Phillips, talk at DPF 2006, Honolulu, Hawaii, October 2006,
{\tt http://www.phys.hawaii.edu/indico/contributionDisplay.py?\\contribId=454\&amp;sessionId=186\&amp;confId=3}.

\bibitem{tevbo}
J.~Nachtman, talk at Fermilab Workshop on the Hunt for Dark Matter, May 2007,
{\tt http://conferences.fnal.gov/dmwksp/Talks/nachtman\_dm\_may07.pdf}.

\bibitem{isa}
  F.~E.~Paige, S.~D.~Protopopescu, H.~Baer and X.~Tata,
  arXiv:hep-ph/0312045.

\bibitem{nusf}
  C.~F.~Kolda and S.~P.~Martin,
  Phys.\ Rev.\  D {\bf 53}, 3871 (1996)
  [arXiv:hep-ph/9503445].

\bibitem{sandre}
  Y.~Santoso,
  arXiv:0709.3952 [hep-ph].

\bibitem{levey}
  S.~J.~J.~Gates and O.~Lebedev,
  Phys.\ Lett.\  B {\bf 477}, 216 (2000)
  [arXiv:hep-ph/9912362].

\bibitem{Arvanitaki:2005nq}
  A.~Arvanitaki, S.~Dimopoulos, A.~Pierce, S.~Rajendran and J.~G.~Wacker,
  Phys.\ Rev.\  D {\bf 76}, 055007 (2007)
  [arXiv:hep-ph/0506242].

\bibitem{Mohapatra:1999gg}
  R.~N.~Mohapatra, F.~I.~Olness, R.~Stroynowski and V.~L.~Teplitz,
  Phys.\ Rev.\  D {\bf 60}, 115013 (1999)
  [arXiv:hep-ph/9906421].


\bibitem{shadrons}
  A.~C.~Kraan,
  Eur.\ Phys.\ J.\  C {\bf 37}, 91 (2004)
  [arXiv:hep-ex/0404001].
  R.~Mackeprang and A.~Rizzi,
  Eur.\ Phys.\ J.\  C {\bf 50}, 353 (2007)
  [arXiv:hep-ph/0612161].

\bibitem{calc}
  A.~Pukhov,
  arXiv:hep-ph/0412191.

\bibitem{pythia} 
  T.~Sjostrand, S.~Mrenna and P.~Skands,
  JHEP {\bf 0605}, 026 (2006)
  [arXiv:hep-ph/0603175].

\bibitem{parton} 
  H.~L.~Lai {\it et al.},
  Phys.\ Rev.\ D {\bf 51}, 4763 (1995)
  [arXiv:hep-ph/9410404].

\bibitem{been1}
  W.~Beenakker, M.~Kramer, T.~Plehn, M.~Spira and P.~M.~Zerwas,
  Nucl.\ Phys.\  B {\bf 515}, 3 (1998)
  [arXiv:hep-ph/9710451].

\bibitem{been2}
 W.~Beenakker, R.~Hopker, M.~Spira and P.~M.~Zerwas,
  Nucl.\ Phys.\  B {\bf 492}, 51 (1997)
  [arXiv:hep-ph/9610490].

  \bibitem{btag}
  ATLAS: Detector and physics performance technical design report,
  Volume 1,
  p317-346,
  1999.
  CERN-LHCC-99-14

\bibitem{Kribs:2007ac}
  G.~D.~Kribs, E.~Poppitz and N.~Weiner,
  arXiv:0712.2039 [hep-ph].

\bibitem{BM}
  J.~R.~Ellis, C.~Kounnas and D.~V.~Nanopoulos,
  Nucl.\ Phys.\  B {\bf 247}, 373 (1984)
;  J.~R.~Ellis, K.~Enqvist, D.~V.~Nanopoulos and K.~Tamvakis,
  Phys.\ Lett.\  B {\bf 155}, 381 (1985);
M.~Drees,
  Phys.\ Lett.\  B {\bf 158}, 409 (1985);
 A.~Corsetti and P.~Nath,
  Phys.\ Rev.\  D {\bf 64}, 125010 (2001)
  [arXiv:hep-ph/0003186];
 N.~Chamoun, C.~S.~Huang, C.~Liu and X.~H.~Wu,
  Nucl.\ Phys.\  B {\bf 624}, 81 (2002)
  [arXiv:hep-ph/0110332];
  U.~Chattopadhyay, D.~Choudhury and D.~Das,
  Phys.\ Rev.\  D {\bf 72}, 095015 (2005)
  [arXiv:hep-ph/0509228];
  S.~Bhattacharya, A.~Datta and B.~Mukhopadhyaya,
  JHEP {\bf 0710}, 080 (2007)
  [arXiv:0708.2427 [hep-ph]].  













\end{thebibliography}
\end{document}